\documentclass[vecphys]{svmult}

\usepackage{makeidx}         % allows index generation
\usepackage{graphicx}        % standard LaTeX graphics tool
\usepackage{natbib}          %
\usepackage{multicol}        % used for the two-column index
\usepackage[bottom]{footmisc}% places footnotes at page bottom
\makeindex             % used for the subject index
\newcommand{\ha}{H$\alpha$ }
\newcommand{\msun}{$M\odot$}

\begin{document}

%\title*{Contribution Title}
\title*{Are the compact star clusters in M82 evolving towards globular 
clusters?}
\titlerunning{Star clusters in M82}
\author{Y.D. Mayya, D. Rosa-Gonz\'alez, L. Rodr\'{\i}guez, L. Carrasco, 
R. Romano \and  A. Luna}
\authorrunning{Mayya et al.}
\institute{Instituto Nacional de Astrof\'{\i}sica, Optica y Electronica, 
Luis Enrique Erro 1, Tonantzintla, C.P. 72840, Puebla, Mexico
\texttt{ydm@inaoep.mx}}
%\author{Name of Author\inst{1}\and Name of Author\inst{2}}
%\institute{Name and Address of your Institute
%\texttt{name@email.address}
%\and Name and Address of your Institute \texttt{name@email.address}}
\maketitle

\begin{abstract}
\footnote{To appear in proceedings of the Puerto Vallarta Conference on ``New 
Quests in Stellar Astrophysics II: Ultraviolet Properties of Evolved Stellar 
Populations'' eds. M. Chavez, E. Bertone, D. Rosa-Gonzalez \& 
L. H. Rodriguez-Merino, Springer, ASSP series.}
%L. H. Rodriguez-Merino and will be published by Springer in the 
%Astrophysics and Space Science Proceedings series.}
%
Recent HST/ACS images of M82 covering the entire galaxy have been used to
detect star clusters. The galaxy is known to contain a young population 
(age $<10^7$~yr) in its starburst nucleus, surrounded by a post-starburst 
disk of age $<10^9$~yr. We detect more than 650 star clusters in this galaxy, 
nearly 400 of them in the post-starburst disk. These data have been used to 
derive the luminosity, mass and size functions separately for the young 
nuclear, and intermediate-age disk clusters. In this contribution, we discuss 
the evolutionary status of these clusters, especially, on the chances of some 
of these clusters surviving to become old globular clusters.
\end{abstract}

\section{Introduction}
\label{sec:1}

Super star clusters (SSCs) and globular clusters (GCs) represent the
youngest and the oldest stellar aggregates known in the Universe.
The environments in which these two kinds of clusters are found are
vastly different --- SSCs are found in violent star-forming regions, 
whereas GCs are found in the halos of galaxies. Yet, the similarity in 
their compactness and mass, is a reason compelling enough to think of 
an evolutionary connection between them. The growing popularity of the 
hierarchical model of galaxy formation in the years following the 
discovery of SSCs, and the possibility of observing the epochs of galaxy (and 
GC) formation at high redshifts, have also generated interest in looking for 
a common origin for these two seemingly different classes of clusters.

In order to investigate the relation between the two types of clusters, it
is important to analyze the survival of SSCs for a Hubble time.
Star clusters are vulnerable to a variety of disruption processes that 
operate on three different timescales \citep[see][for more details]
{Fal01, Mai04, deG07}. On short timescales ($t\sim10^7$~yr), the exploding 
supernovae and the resulting superwinds are responsible for cluster expansion
and disruption, a process popularly dubbed as infant mortality. 
On intermediate timescales ($10^7<t<{\rm few} \times 10^8$~yr),
the mass-loss from evolving stars leads to the disruption of the clusters.
On even longer timescales ($t> {\rm few} \times 10^8$~yr),
stellar dynamical processes, especially evaporation due to two-body scattering,
and tidal effects on a cluster as it orbits around the galaxy, known as
gravitational shocks, come into play in the removal of stellar mass from
clusters. The GCs represent those objects that have survived all these 
processes, whereas young SSCs are just experiencing them.
Intermediate age SSCs are the ideal objects to investigate 
the influence of disruption processes on the survival of star clusters.
Almost all the star formation in the disk of M82 took place in a violent 
disk-wide burst around 100--500~Myr ago, following the interaction of M82 
with the members of M81 group \citep{May06}. Cluster formation is known to 
be efficient during the burst phase of star formation \citep{Bas05}, and 
hence we expect large number of clusters of intermediate age ($\sim100$~Myr) 
in its disk. Hence, M82 offers an excellent opportunity to assess the 
evolutionary effects on the survival of star clusters, and to look for a 
possible evolutionary connection between the SSCs and GCs.

\section{Data Extraction, Source Selection, and Simulations}

The observational data used in this work consisted of images in F435W ($B$), 
F555W ($V$) and F814W ($I$) filters, that were obtained by the Hubble Heritage 
Team \citep{Mut07} using the ACS/WFC instrument aborad the Hubble Space 
Telescope (HST). Bias, dark, and flat-field corrections were carried out 
using the standard pipeline process by the Heritage Team. The final reduced 
science quality images cover the entire optical disk of the galaxy with a 
spatial sampling of 0.05~arcsec\,pixel$^{-1}$, which corresponds to 
0.88~parsec\,pixel$^{-1}$ at M82's distance of 3.63 Mpc \citep{Fre94}.
The point sources have a size distribution that peaks at a Full Width at Half
Maximum (FWHM) of 2.1~pixels, with the tail of the distribution extending to 
3.0~pixels (or 2.6 parsec). Very few clusters are expected to have 
sizes smaller than 3~parsec, and hence clusters can be distinguished from the 
stars on these images.
A circle of 500~pixels (450~pc) radius is used to separate the nuclear
region from the disk. The clusters inside this radius are associated
with strong H$\alpha$ emitting complexes, and hence are younger
than 10~Myr \citep{Mel05}. On the other hand, the disk outside the 450~pc
radius shows characteristic signatures of post-starburst conditions,
with hardly any \ha\  emission.

We used SExtractor \citep{Ber96} independently on the $B$, $V$, and $I$-band 
images to construct an unbiased sample of cluster candidates. A source 
having a FWHM $>3$~pixels and an area of at least 50 adjacent pixels, each of 
S/N $>5$ is considered a cluster candidate. All the bright sources satisfying
these criteria are genuine clusters, but at fainter magnitudes majority of 
the candidate sources lack the symmetry expected for a physical cluster.
These are found to be artificial extended sources formed due to the 
superposition of stars in this nearly edge-on galaxy. These artificial
sources most often are elongated, and are rejected automatically
from the sample using the ellipticity parameter of SExtractor.
Cluster candidates in each filter were then combined, the common sources 
being counted only once. The resulting list contains
653 clusters, 260 of them belonging to the nuclear region.
For all the sources in the final list, aperture photometry is carried out 
in all the three bands. The FWHM calculated by SExtractor is used as a 
measure of the size of the clusters.

The observed cluster luminosity function (LF) follows a power-law at the 
bright end, turning over sharply at faint magnitudes. Similarly, the size
distribution function peaks at a characteristic value of FWHM$\sim$10~pixels.
%In order to investigate whether observational biases are responsible
%for these turn-overs, we carried out detailed Monte Carlo simulations.
%
Monte Carlo simulations were carried out to check the effect of incompleteness 
of cluster detection on the observed functional forms.
In the simulations, each
cluster is assumed to be round and to have a Gaussian intensity profile
of a given FWHM. A power-law distribution function is used to assign a 
luminosity to each cluster. Two separate simulations are done, one in which 
a cluster is assigned a size based on a power-law size 
distribution function, and the other based on a log-normal function.
The simulated luminosity function resembles very much the observed one,
implying that the observed turn-over of the luminosity function is due
to incompleteness at the faint end and not intrinsic to the cluster
population. Hence, the turn-over in the luminosity function, if any,
would correspond to a magnitude fainter than $B=22$ mag.
On the other hand, the observed size function points to an intrinsically 
log-normal size distribution, rather than a power-law function.
A more detailed description of the selection process, observed luminosity and
size functions, and the Monte Carlo simulations can be found in \citet{May07}.

\section{Physical Parameters of Clusters}

We analyzed the color and magnitude of the individual clusters to obtain their
reddening and mass, making use of solar metallicity Single Stellar Population
(SSP) models of \citet{Gir02}. These authors provide the evolutionary data on
colors and magnitudes for the instrumental HST/ACS filters,
a fact that enables us a direct comparison with the observed data.
The \citet{Kro01} initial mass function (IMF) in its corrected version 
has been used. It has nearly a Salpeter slope (2.30 instead of 2.35) for 
all masses higher than 1~\msun. The derived masses depend on the assumption
of the lower cut-off mass of the IMF. In the case of standard Kroupa's IMF,
the derived masses would be around 2.5 times higher. 

\subsection{The Ages of Clusters}

Colors obtained by the combination of the three filters that we used suffer
from age-reddening degeneracy, and hence it was necessary to assume one of
the quantities to obtain the other. We found that the observed range in 
colors is too large to be explained by evolutionary effects, even for 
stellar populations as old as 10~Gyr. On the other hand, age of the principal
stellar populations in the nuclear region of M82 is determined in innumerable 
studies \citep[and references therein]{Rie93, For03}, and it is 
found to be $<10$~Myr.
Based on these studies, we adopt an age of 8~Myr for the nuclear clusters.
Most of the disk stars in M82 were formed in a violent burst around 500~Myr
ago. Ages of those clusters for which spectroscopic data are available 
\citep{Smi06} lie in the range between 50--500~Myr, suggesting that the clusters 
are formed during or immediately after the disk-wide star formation epoch.

\begin{figure}
\centering
\includegraphics[height=12cm]{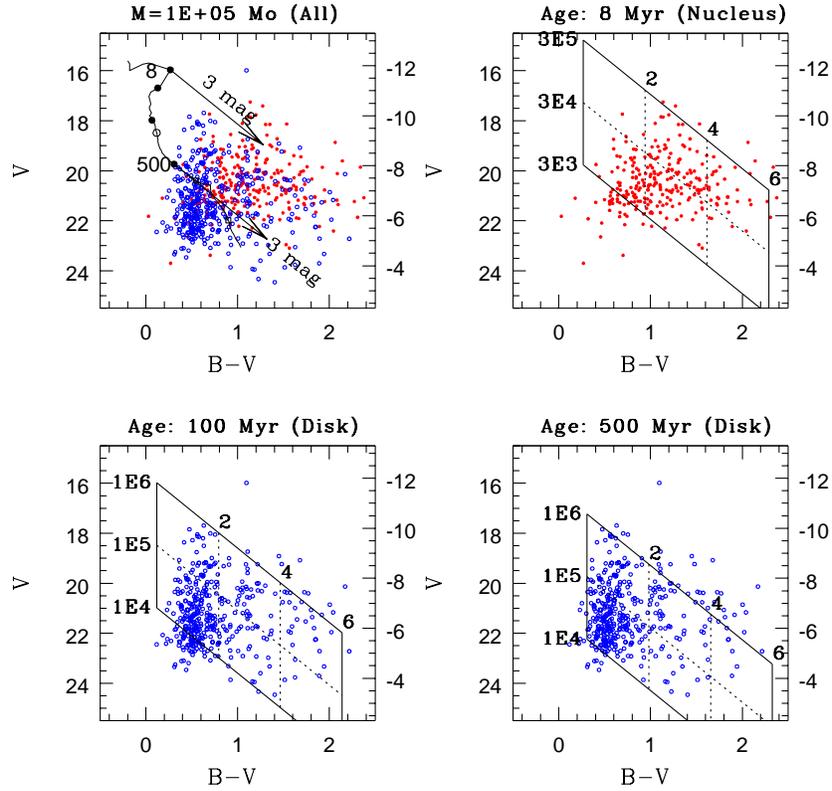}
\caption{Observed color-magnitude diagrams (CMDs) for the nuclear (filled
circles) and disk clusters, in M82. ({\it Top left}) Evolutionary track 
for an SSP of a cluster mass of $10^5$~$M\odot$ is superposed. 
Two vectors, placed at 8~Myr and 500~Myr, show the location of the
track reddened by $A_v=3$~mag.   
In the top-right panel, we show the CMD for the nuclear clusters only.
The locations of an 8~Myr SSP for a range of cluster masses and visual
extinctions are shown by the superposed grid. Mass varies vertically
along the grid (in solar units), whereas the visual extinction (in magnitude)
varies along the diagonal axis.
In the bottom panels, we show a similar diagram for the disk clusters,
with the superposed grids corresponding to fixed ages of 100~Myr (left)
and 500~Myr (right). In all the panels, tick mark values of the
right-vertical axis correspond to the absolute magnitude in the $V$-band.
\label{fig_cmds}
}
\end{figure}

\subsection{Color-Magnitude Diagrams}

From the assumed ages (8~Myr for the nuclear clusters, and 50--500~Myr
for the disk clusters), and the very likely hypothesis that the extinction
is the main cause of the dispersion in the observed colors, we can estimate
the masses of the clusters. The method we have followed is illustrated
in Figure~\ref{fig_cmds}. For a given position in the Color Magnitude
Diagram (CMD), we derived the extinction by comparing
the observed colors with those of the SSP. Once the extinction is
determined, we calculate the mass using the extinction-corrected
luminosity and the mass-to-light ratio of such SSP. The disk masses are
derived assuming an age of 100~Myr. The mass estimates would be higher by
a factor of 3.2, if the clusters are as old as the stellar disk (500~Myr).
On the other hand, if the clusters are as young as 50~Myr, the masses
would be lower by a factor of 1.6. The distribution of the derived visual
extinction values is peaked at $\sim1$~mag for the disk clusters, whereas 
it is flat between 1--4~mag for the nuclear clusters.

\begin{figure}
\centering
\includegraphics[height=5.5cm]{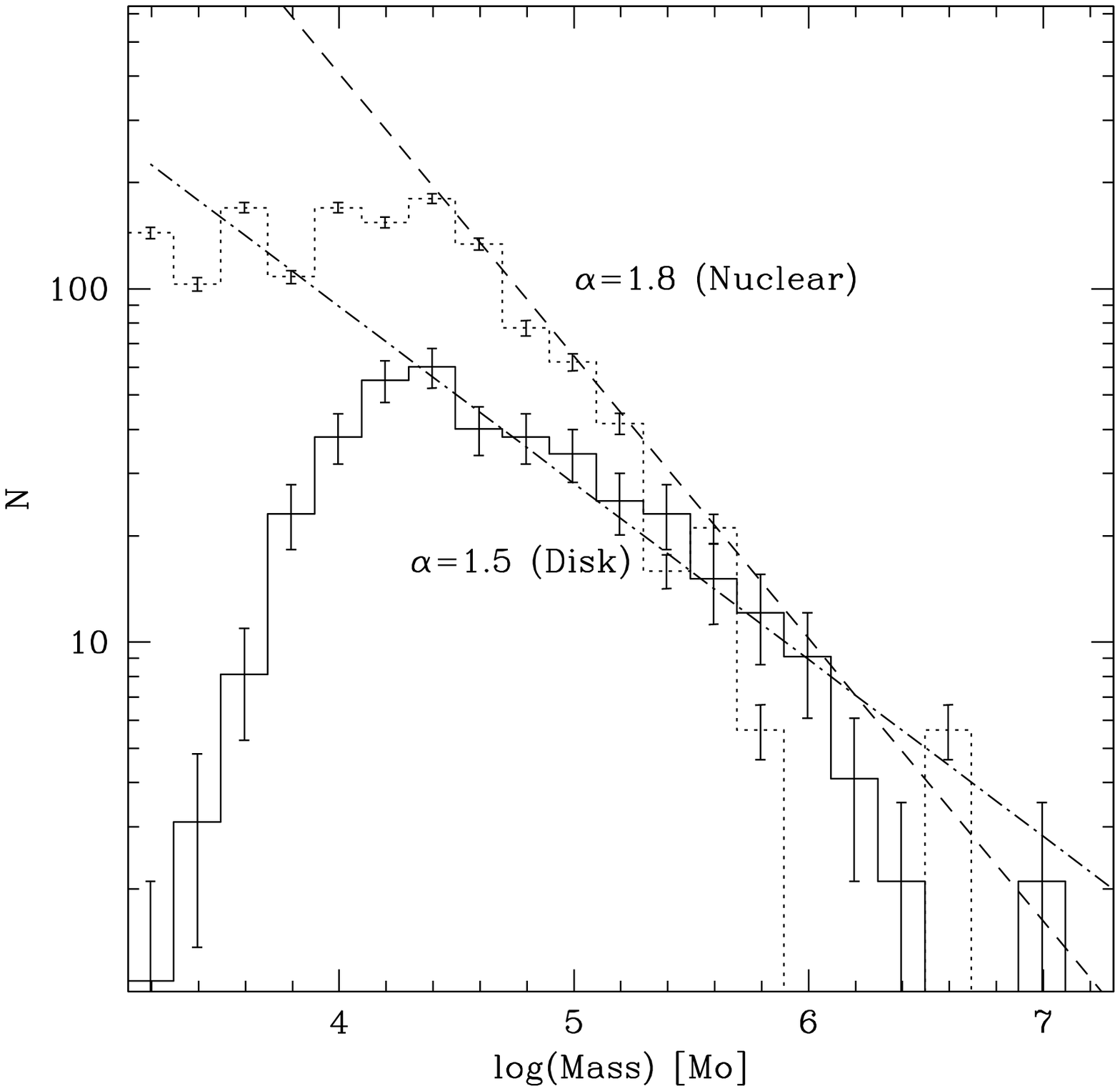}
\includegraphics[height=5.5cm]{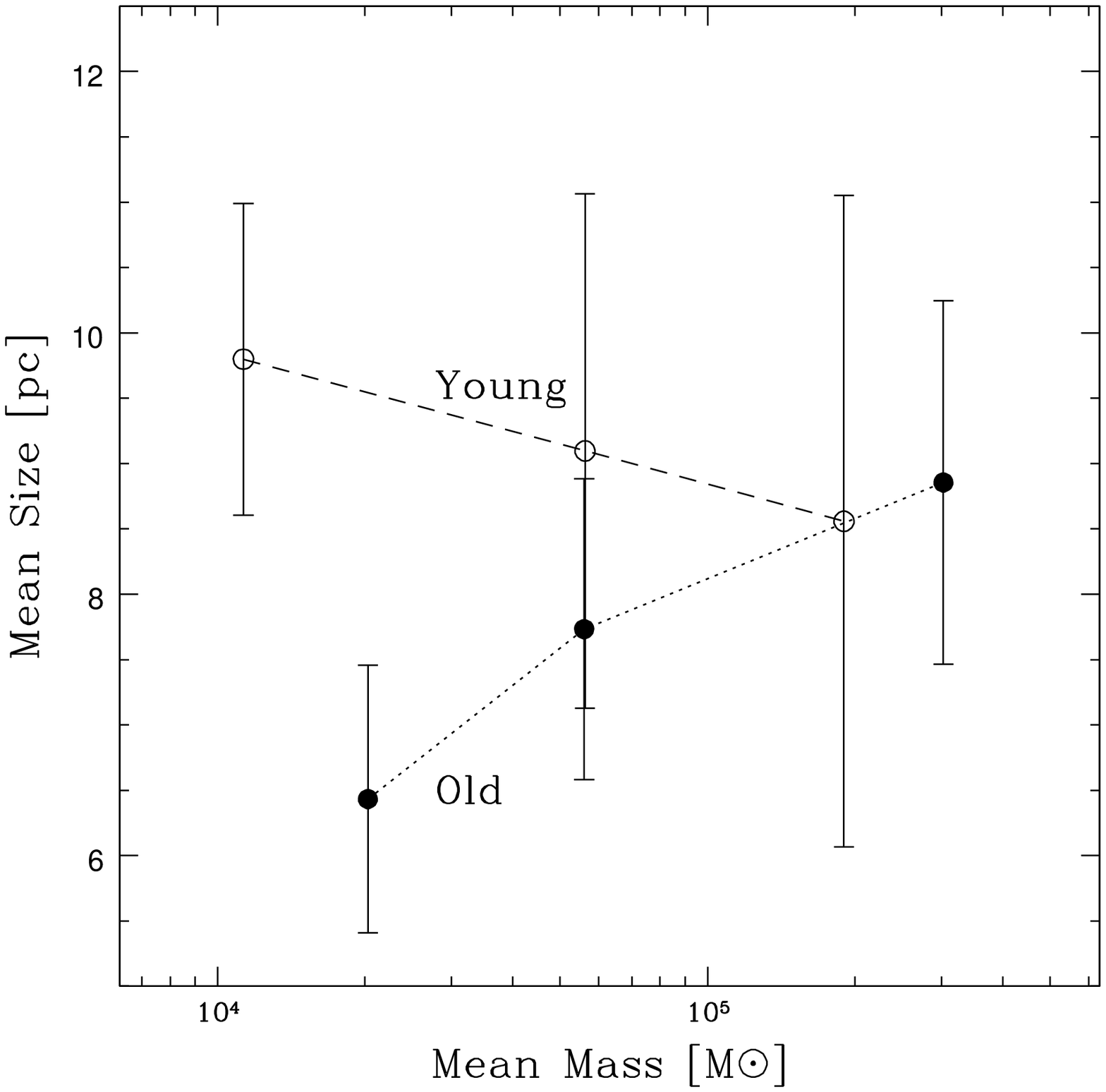} 
\caption{(Left)
Mass functions for the nuclear (dotted line) and disk (solid line)
cluster samples. Both the samples follow a power-law distribution between
$2\times10^4$\msun\ and $10^6$\msun. The best-fit indices in this mass
range are indicated.
(Right) Mean size (FWHM) of the clusters as a function of mean mass for
three mass bins for the nuclear (young) and disk (old) samples.
The error bar denotes the rms dispersion about the mean value.
High mass clusters have similar mean sizes irrespective of their
evolutionary status. On the other hand, mean size of the low-mass clusters
clearly decreases as they become older. Among the young clusters, low-mass
ones are more extended than higher mass ones.
%\label{fig_stats}
\label{fig_mass_fun}
}
\end{figure}
               
\subsection{Mass Distribution Function}

The determination of the cluster masses for our complete sample enables us to
derive the present-day Cluster Mass Function (CMF). In the left panel of
Figure~\ref{fig_mass_fun}, we plot the CMF separately for the nuclear and 
disk clusters. The nuclear CMF is scaled up to match the disk CMF 
at $1.5\times10^6$~\msun. Poissonian error bars are indicated. The 
distribution for both samples follows a power-law over
almost two orders of magnitude in mass for cluster masses
above $\sim2\times10^4$~\msun. However, the power-law index for the
disk and nuclear cluster populations shows a clear
difference, $\alpha=1.8\pm0.1$ for the nuclear clusters, and $1.5\pm0.1$
for the disk population. Studies of young star clusters in nearby galaxies 
yield a value of $\alpha$ close to 2.0 \citep{deG03}. Hence, $\alpha=2.0$ 
can be considered as the expected slope of the initial CMF.

In general, the cluster size distribution function (CSF) for the 
nuclear and disk clusters follow a log-normal form. However, the mean, 
as well as the maximum cluster sizes are systematically smaller for the 
lower mass bins.
This tendency is illustrated in the right panel of Figure~\ref{fig_mass_fun}, 
where the mean
cluster size for each mass bin has been plotted against the mean mass of
clusters in that bin, for the young and old ones, separately. For the highest
mass bin, the mean sizes of the young and old clusters are similar. The
mean size decreases systematically with decreasing cluster mass for the
old clusters, whereas the inverse is true for the young clusters.

\section{On the Survival Chances of Star Clusters in M82}

The observed differences in the CSF for young and old clusters are
consistent with the expected evolutionary effects. Both the disruption
of the loose OB associations and the dynamical trend towards relaxation
would diminish the number of large low-mass systems. Thus, the destruction
process is both mass and size dependent, with the most extended clusters
in each mass bin being the most vulnerable to disruption. All clusters of 
masses higher than $10^5$~\msun\ are still surviving $\sim10^8$~yr after their
formation. In this section, we discuss these observational
results in the context of theoretical models of cluster disruption, and
their possible survival to become globular clusters.

At early times, disruption is caused mainly due to the expulsion 
of the intra-cluster gas through supernova explosions. This process is
ineffective once all the high mass stars in the cluster die, which happens
in around $\sim30$~Myr. Hence, the observed disk clusters have survived this
early mechanism of disruption.
On intermediate timescales ($10^7<t<{\rm few} \times 10^8$~yr),
the mass-loss from evolving stars leads to a decrease in the cluster mass
from its initial value. Clusters can loose as much as 30\% of their stellar
mass during their evolution. The decreased cluster mass can result in the
expansion of the cluster, finally leading to its disruption. However, this
process of disruption is equally effective for high and low mass clusters,
and hence a change in the slope of mass function is not expected. The
observed flattening of the mass function at older ages, suggests that the
cluster disruption process that is active in M82 selectively destroys
low-mass clusters. The tidal effect experienced by the clusters as they move 
in the gravitational field of the parent galaxy is one such process.
According to \citet{Fal01}, this process becomes important
after $\sim300$~Myr in normal galaxies. However, in the case of M82,
\citet{deG05} have estimated a disruption timescale as short as 30~Myr
for a cluster of mass $10^4$~\msun\  at 1~kpc away from the center, with
a dependence on mass that varies as $M^{0.6}$. The short timescale in 
M82 implies that the surviving clusters in the disk are presently experiencing
the dynamical processes of cluster disruption. 

\begin{figure}
\centering
\includegraphics[height=11cm]{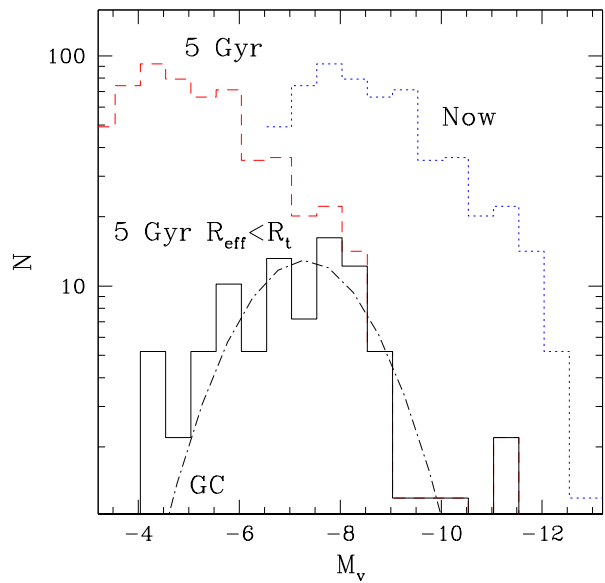}
\caption{Present (dotted histogram) and future luminosity functions of M82
star clusters with (solid histogram) and without (dashed histogram) 
taking into account dynamical effects of evolution at an age of 5~Gyr. 
The log-normal 
function representing the luminosity function of the Galactic Globular 
Clusters is shown by the dot-dashed curve.
\label{fig_lumfun}
}
\end{figure}

If the trend of selective disruption of loose clusters continues,
how many of the present clusters will survive for a Hubble time? 
Can the LF of the surviving clusters look like that of the Galactic GCs?
In Figure~\ref{fig_lumfun}, we show the evolutionary effects on the LF 
of the M82 clusters. The histogram with dashed lines shows the LF 
considering the photometric evolution of the clusters for 5~Gyr, whereas 
the solid histogram shows the same, but after taking into account the dynamical 
effects as well. The latter is implemented in a simplistic way, by
imposing the condition that for the clusters to survive the dynamical
effects, their half-light radius, $R_{\rm eff}$, should be smaller than 
the tidal radius, $R_{\rm t}$, for that cluster. For a cluster of 
mass $M_{\rm C}$ at galactocentric radius $R_{\rm G}$, $R_{\rm t}$ 
is given by the expression \citep{Spi87}, \\
$R_{\rm t} = \left({M_{\rm C}\over{2M_{\rm G}}}\right)^{1/3} R_{\rm G}$,\\
where $M_{\rm G}$ is the mass of the parent galaxy, which for M82 is somewhat
uncertain due to the difficulty in interpreting uniquely the observed gas
velocity fields in this disturbed galaxy, and the best estimate is 
$M_{\rm G}=10^{10}$\msun \citep{Sof98}. The $R_{\rm t}$ values calculated
using the currently observed galactocentric distances indicate that most of
the nuclear clusters, and very few of the disk clusters, will be destroyed.
Thus, if the clusters are in circular orbits, the LF 
will practically retain its present power-law form.  
However, the stellar orbits in the central bar of M82 are known to be highly
elliptical \citep{Gre02}, which implies that the galactocentric distance 
of a cluster will 
change with time. The disruption of a cluster depends on the net 
tidal force received by its stars as it orbits the galaxy during its 
lifetime. We found that for an assumed $R_{\rm G}=350$~pc, the future LF of 
M82 will resemble that of the Galactic GCs.
Even in this extreme case, 85 clusters will survive, as compared to 
the 146 GCs in the Milky Way. The number of GCs in a galaxy scale with the 
mass of the parent galaxy, and considering that M82 is an order of magnitude 
less massive than the Milky Way, only $\sim15$~GCs are expected to present 
in M82. Thus, number of clusters that will survive represent an
over-abundance by a factor of 5--6. For ages older than this, the 
distribution is similar except that the peak of the distribution shifts 
to $\sim$0.5~mag fainter. Thus, the compact star clusters in M82 will evolve
into GCs.

\section{Conclusions}

Luminosity and Mass functions of star clusters in M82 follow power-law
functions, with the power law index showing a tendency for flattening
of the profile with age. In other words, there is a deficiency of 
low-mass clusters among the older clusters. We also find the
mean size of the older clusters to be smaller as compared to the younger
clusters for masses $<10^5$~\msun. These two results together imply the
selective destruction of loose clusters. The tidal forces experienced
by the clusters as they orbit around the galaxy lead to exactly such 
a destruction process. If this process continues in M82, 
the LF of surviving clusters can mimic the presently
observed LF of the Galactic GCs, provided the clusters move around the
galaxy in highly elliptical orbits, with perigalactic distance as small
as 350~pc. The resulting LF contains 85 clusters with the function peaking 
at the same luminosity as for the Galactic GCs at 5~Gyr age, and fainter by 
 $\sim$0.5~mag at 10~Gyr. 
On the other hand, if the clusters move in nearly circular orbits, the LF
will retain the power-law form, with the number of surviving clusters even
higher.

\vspace*{0.3cm}
This work is partly supported by CONACyT (Mexico) research grants
42609-F and 49942-F.
We would like to thank the Hubble Heritage Team at the Space Telescope
Science Institute for making the reduced fits files available to us.
%\begin{thebibliography} %{99.}

\end{document}